\documentclass[12pt,a4]{article}
\usepackage{epsfig,amsmath,amsfonts}
\begin{document}
\input{epsf}

\newcommand{\be}{\begin{equation}}
\newcommand{\ee}{\end{equation}}
\newcommand{\bea}{\begin{eqnarray}}
\newcommand{\eea}{\end{eqnarray}}
\newcommand{\rf}[1]{(\ref{#1})}
\newcommand{\pa}{\partial}
\newcommand{\nn}{\nonumber}
\newcommand{\e}{\mbox{e}}
\renewcommand{\d}{\mbox{d}}
\newcommand{\g}{\gamma}
\renewcommand{\l}{\lambda}
\renewcommand{\L}{\Lambda}
\renewcommand{\b}{\beta}
\renewcommand{\a}{\alpha}
\newcommand{\n}{\nu}
\newcommand{\m}{\mu}
\newcommand{\Tr}{\mbox{Tr}}
\newcommand{\E}{\mbox{E(q)}}
\newcommand{\Ee}{\mbox{E}}
\newcommand{\K}{\mbox{K(q)}}
\newcommand{\Kk}{\mbox{K}}
\newcommand{\ep}{\varepsilon}
\newcommand{\om}{\omega}
\newcommand{\del}{\delta}
\newcommand{\Del}{\Delta}
\newcommand{\sg}{\sigma}
\newcommand{\vph}{\varphi}
\newcommand{\sn}{\mbox{sn}}
\newcommand{\dn}{\mbox{dn}}
\newcommand{\cn}{\mbox{cn}}
\newcommand{\CA}{\mathcal{A}}
\newcommand{\oh}{\frac{1}{2}}
\newcommand{\oq}{\frac{1}{4}}
\newcommand{\dg}{\dagger}
\newcommand{\R}{\mathbb{R}}
\newcommand{\ra}{\rightarrow}
\newcommand{\la}{\left\langle}
\newcommand{\prt}{\partial}
\newcommand{\mi}{\!-\!}
\newcommand{\equ}{\!=\!}
\newcommand{\pl}{\!+\!}
\newcommand{\CN}{\mathcal{N}}
\newcommand{\cD}{{\cal D}}
\newcommand{\cS}{{\cal S}}
\newcommand{\cM}{{\cal M}}
\newcommand{\cK}{{\cal K}}
\newcommand{\cT}{{\cal T}}
\newcommand{\cN}{{\cal N}}
\newcommand{\cL}{{\cal L}}
\newcommand{\cO}{{\cal O}}
\newcommand{\cR}{{\cal R}}
\newcommand{\CH}{\mathcal{H}}
\newcommand{\CL}{\mathcal{L}}
\newcommand{\CO}{\mathcal{O}}
\newcommand{\CI}{\mathcal{I}}
\newcommand{\CT}{\mathcal{T}}
\newcommand{\CS}{\mathcal{S}}
\newcommand{\CM}{\mathcal{M}}
\newcommand{\CQ}{\mathcal{Q}}
\newcommand{\CE}{\mathcal{E}}
\newcommand{\CB}{\mathcal{B}}
\newcommand{\tF}{{\tilde{F}}}
\newcommand{\tL}{{\tilde{\L}}}
\newcommand{\tX}{{\tilde{X}}}
\newcommand{\tY}{{\tilde{Y}}}
\newcommand{\tZ}{{\tilde{Z}}}
\newcommand{\ty}{{\tilde{y}}}
\newcommand{\tz}{{\tilde{z}}}
\newcommand{\tg}{{\tilde{g}}}
\newcommand{\tG}{{\tilde{G}}}
\newcommand{\tH}{{\tilde{H}}}
\newcommand{\tT}{{\tilde{T}}}

\newcommand{\tSL}{\sqrt{\tL}}
\newcommand{\FL}{\L^{1/4}}
\newcommand{\bZ}{{\bar{Z}}}
\newcommand{\bX}{{\bar{X}}}

\newcommand{\remark}[1]{{\renewcommand{\bfdefault}{b}\textbf{\mathversion{bold}#1}}}
\newcommand{\gLs}{(\gamma L)^2}
\newcommand{\doublet}[2]{\left(\begin{array}{c}#1\\#2\end{array}\right)}
\newcommand{\twobytwo}[4]{\left(\begin{array}{cc} #1&#2\\#3&#4\end{array}\right)}
\newcommand{\threebythree}[9]{\left(\begin{array}{ccc} #1&#2&#3\\#4&#5&#6\\#7&#8&#9\\\end{array}\right)}
\newcommand{\iq}{q^{-1}}
\newcommand{\half}{\frac12}
\newcommand{\I}{{\mathbb{I}}}
\newcommand{\Rhat}{\hat{\mathrm{R}}}
\newcommand{\suq}{\mathrm{su}_q(3)}
\newcommand{\Rh}[4]{\hat{R}^{#1\phantom{#2}#3}_{\phantom{#1}#2\phantom{#3}#4}}
\newcommand{\Rt}[4]{\widetilde{R}^{#1\phantom{#2}#3}_{\phantom{#1}#2\phantom{#3}#4}}
\newcommand{\Urm}{\mathrm{U}}
\newcommand{\SU}{\mathrm{SU}}
\newcommand{\SO}{\mathrm{SO}}
\newcommand{\Sp}{\mathrm{Sp}}
\newcommand{\su}{\mathrm{su}}
\newcommand{\SL}{\mathrm{SL}}
\newcommand{\GL}{\mathrm{GL}}
\newcommand{\tq}[2]{\mathrm{t}^{#1}_{\phantom{#1}\!#2}}
\newcommand{\sq}[2]{\mathrm{s}^{#1}_{\phantom{#1}\!#2}}
\newcommand{\Wcal}{\mathcal{W}}
\newcommand{\mparagraph}[1]{\paragraph{#1}\mbox{}\vspace{.04cm}}
\newcommand{\Phibar}{\overline{\Phi}}
\newcommand{\Zset}{\mathbb{Z}}
\newcommand{\Ncal}{\mathcal{N}}
\newcommand{\Fcal}{\mathcal{F}}
\newcommand{\Hcal}{\mathcal{H}}
\newcommand{\Acal}{\mathcal{A}}
\newcommand{\Ucal}{\mathcal{U}}
\newcommand{\Vcal}{\mathcal{V}}
\newcommand{\Mcal}{\mathcal{M}}
\newcommand{\Rcal}{\mathcal{R}}
\newcommand{\AdS}{\mathrm{AdS}}
\newcommand{\Srm}{\mathrm{S}}
\newcommand{\Cset}{{\,\,{{{^{_{\pmb{\mid}}}}\kern-.47em{\mathrm C}}}}}
\newcommand{\Supertwistor}{\Cset \mathrm{P}^{3|4}}
\newcommand{\ket}[1]{\left|#1\right.\rangle}
\newcommand{\qb}{{\bar{q}}}
\newcommand{\hb}{{\bar{h}}}
\newcommand{\diff}{\mathrm{d}}
\newcommand{\RP}{\mathbb{R}\mathrm{P}}

\thispagestyle{empty}

\begin{center}

{\large \bf {On the spectral problem of ${\cal N}=4$ SYM with 
\\ orthogonal or symplectic gauge group}}

\vspace*{26pt}

{\sl Pawel Caputa$\, ^a$}, {\sl Charlotte Kristjansen$\, ^b$} and
{\sl Konstantinos Zoubos$\, ^b$}

\vspace{10pt}
\vspace{10pt}

$^a$The Niels Bohr International Academy, \\
The Niels Bohr Institute, Copenhagen University\\
Blegdamsvej 17, DK-2100 Copenhagen \O , Denmark\\\vspace{.4cm}
$^b$ The Niels Bohr Institute, Copenhagen University\\
Blegdamsvej 17, DK-2100 Copenhagen \O , Denmark\\\vspace{.4cm}
{\tt\small caputa@nbi.dk, kristjan@nbi.dk, kzoubos@nbi.dk}
\vspace{10pt}

\end{center}

\begin{abstract}
\noindent
We study the spectral problem of ${\cal N}=4$ SYM with gauge group $SO(N)$
and $Sp(N)$. At the planar level, the difference to the case of
gauge group $SU(N)$ is only due to certain states being projected out, however
at the non-planar level novel effects appear: While 
$\frac{1}{N}$-corrections in the $SU(N)$ case are always associated with
splitting and joining of spin chains, this is not so for $SO(N)$ and
$Sp(N)$. Here the leading $\frac{1}{N}$-corrections, which are due
to non-orientable Feynman diagrams in the field theory, 
originate from a term in the dilatation
operator which acts inside a single spin chain. This makes it possible
to test for integrability of the leading $\frac{1}{N}$-corrections 
by standard (Bethe ansatz) means and we carry out various such tests. 
For orthogonal and symplectic gauge group the dual string theory lives
on the orientifold $\AdS_5\times \mathbb{R}\mbox{P}^5$. We discuss various issues 
related to semi-classical strings on this background.

\end{abstract}

\newpage

\setcounter{page}{1}

\newcommand{\ft}[2]{{\textstyle\frac{#1}{#2}}}
\newcommand{\ii}{\mathrm{i}}
\newcommand{\dd}{{\mathrm{d}}}
\newcommand{\nnb}{\nonumber}

\section{Introduction}

 Whereas the planar spectral problem of ${\cal N}=4$ SYM seems to be 
close to 
resolution~\cite{Minahan:2002ve,Beisert:2003tq,Beisert:2005fw,
Beisert:2006ez,Beisert:2006ib,Ambjorn:2005wa,Gromov:2009tv,Bombardelli:2009ns,
Arutyunov:2009ur}, 
much less has been achieved in the non-planar case. Non-planar 
corrections, when studied perturbatively in $\frac{1}{N}$, lead to a 
breakdown of the spin chain picture which was the key to the progress at
the planar level.
More precisely, $\frac{1}{N}$-corrections to the dilatation generator lead
to interactions which split and join spin chains~\cite{Beisert:2002ff}. 
This enormously enlarges the Hilbert space of states and, furthermore, 
implies that excitations on different chains can interact, rendering the 
standard tools of integrable spin chains inapplicable and leaving little hope 
for the existence of a Bethe ansatz in the usual sense.\footnote{The situation is the 
same in the three--dimensional ABJM and ABJ theories \cite{Kristjansen:2008ib,Caputa:2009ug}.}

In order to gain further insight into $\frac{1}{N}$-corrections we will study
${\cal N}=4$ SYM with gauge groups $SO(N)$ and $Sp(N)$. At the planar level, 
the only essential difference of these theories from the traditionally studied 
$SU(N)$ case is that certain
states are projected out. However, at the non-planar level new effects
arise. Namely, for orthogonal and symplectic gauge group the leading 
non-planar corrections originate from non-orientable Feynman diagrams with
a single cross-cap~\cite{Cicuta:1982fu}. 
At the level of the dilatation generator 
these leading non-planar corrections are described by an operator which
acts entirely inside a {\it single} spin chain. This implies that restricting
oneself to the leading $\frac{1}{N}$-corrections one does not
face the 
problems mentioned above. The Hilbert space of states remains the same
as on the planar level and all interactions take place inside
a single spin chain. Thus the existence of a usual Bethe ansatz is not a priori
excluded and 
one may test for integrability using standard methods.

In the  AdS/CFT correspondence, 
changing the gauge group on the field theory side translates
into a modification of the background geometry on the string theory side.
 For orthogonal and symplectic gauge groups
the relevant geometry becomes that of the orientifold
$\AdS_5\times \mathbb{R}\mbox{P}^5$ where the case of $Sp(N)$ differs from that of $SO(N)$
by the presence of an additional $B$-field~\cite{Witten:1998xy}.
In the case of ${\cal N}=4$ SYM
with gauge group $SU(N)$ the leading non-planar effects on the string theory
side have their origin in string diagrams of genus one but in the case
of orthogonal and symplectic gauge groups the leading non-planar corrections
should be associated with non-orientable
string worldsheets with a single cross-cap. 
At least naively, it seems
easier to deal with cross-caps than higher genus surfaces so our study
might open new avenues for comparison of gauge and string theory beyond
the planar limit.

Our main focus will be on the gauge theory side where we will study in depth
the one-loop dilatation generator. We start in section~\ref{oneloop}
by explaining the reduction of
the space of states compared to the theory with gauge group $SU(N)$ and 
subsequently write down the one-loop dilatation generator including all
non-planar corrections. In section~\ref{BMN} we determine analytically
the leading $\frac{1}{N}$-correction to the anomalous dimension
of two-excitation states, thereby
providing a prediction for the dual string theory. After that, in 
section~\ref{integrability},
we search for integrability
in the non-planar spectrum in various ways. We look for unexpected degeneracies
and for conserved charges. In addition, 
we put forward various
possible modifications of the planar Bethe equations which would produce
the correct $\frac{1}{N}$-correction for two-excitation states and test 
numerically if these equations also work for higher numbers of excitations.
Unfortunately, the outcome of these tests is negative. In
section~\ref{stringtheory},
we discuss the dual string theory picture and,
in particular, mention 
a number of interesting open problems. Finally,
section~\ref{conclusion} contains our conclusion.

\section{${\cal N}=4$ SYM with gauge group $SO(N)$ \label{oneloop}}

In this section we will study non-planar effects in the spectrum of
${\cal N}=4$ SYM with gauge group $SO(N)$. Before doing so, it is useful to 
briefly recall how this theory arises as a suitable projection of the $SU(N)$
theory. As is well known, in string theory the latter is constructed by 
taking the low-energy limit of a stack of $N$ D3-branes in ten-dimensional 
Minkowski space. The group $SU(N)$ arises because the matrices $\lambda^i_{\;j}$ 
encoding the Chan-Paton factors of the open strings stretching between the D3-branes 
are hermitian. 

In order to obtain an orthogonal gauge group, one performs an orientifold projection
which, on bosonic states, amounts to relating the Chan-Paton matrices to their
transpose matrices as \cite{Gimon:1996rq}
\be
\lambda=-\eta^{-1} \lambda^T \eta
\ee
where $\eta$ is a symmetric matrix which can simply be taken to be unity. The Chan-Paton
matrices are thus restricted to be antisymmetric $N\times N$ matrices, which generate
the adjoint representation of the group $SO(N)$. As explained in \cite{Witten:1998xy}, 
in order to ensure that this procedure does not break  $\Ncal=4$ supersymmetry one
has to combine it with a spacetime identification of the six transverse to the brane 
coordinates $X^i$ as $X^i\ra -X^i$. This procedure leaves us with $\Ncal=4$ SYM with 
gauge group $SO(N)$. 

We will restrict ourselves to considering the $SU(2)$ sub-sector of the theory, consisting of 
multi-trace operators built from two complex fields, say $\phi$ and $Z$, i.e. operators
of the form
\begin{equation}
{\cal O}= \mbox{Tr}(Z\ldots Z  \phi\ldots \phi Z\ldots )
\mbox{Tr}(Z\ldots Z \phi\ldots \phi Z\ldots  )\ldots
\label{operators}
\end{equation}
The adjoint fields $Z$ and $\phi$, being elements of the algebra of $SO(N)$, fulfill
\begin{equation}
\phi^T=-\phi,\hspace{0.7cm} Z^T=-Z.
\end{equation}
The dilatation generator of the $SU(2)$ sub-sector at one and two-loops can formally be 
written in the same way as for the $SU(N)$ case~\cite{Beisert:2003tq}. At one loop order 
it reads\footnote{We chose to keep the normalization of generators $\Tr T^a T^b=\delta^{ab}$
when passing from $SU(N)$ to $SO(N)$.}
\begin{equation}
\hat{D}= -\frac{g_{\mbox{\tiny YM}}^2}{8\pi^2}\,
\mbox{Tr}[\phi,Z][\check{\phi},\check{Z}]
\equiv \frac{g_{\mbox{\tiny YM}}^2}{8\pi^2}\, \hat{H}.
\label{dilop}
\end{equation}
Here $\check{Z}$ is an operator which acts on a field $Z$ by contraction of $SO(N)$ indices, i.e.
\begin{equation} \label{contraction}
\check{Z}_{\alpha\,\beta} Z_{\gamma\epsilon}
=\frac{1}{2}(\delta_{\alpha \epsilon} \delta_{\beta \gamma}-
\delta_{\alpha \gamma} \delta_{\beta \epsilon}),
\end{equation}
and similarly for $\check{\phi}$. 

In the analysis of ${\cal N}=4$ SYM with gauge group $SU(N)$ the concept of
parity played a central role. In a spin chain context, parity is the operation which inverts 
the order of operators inside a given trace, i.e.~\cite{Doikou:1998jh}
\begin{equation}
\hat{P}\mbox{Tr}(X_{i_1} X_{i_2}\ldots X_{i_L})=
\mbox{Tr}(X_{i_L} X_{i_{L-1}}\ldots X_{i_1}).
\end{equation}
Parity commutes with $\hat{H}$ which means that eigenstates of $\hat{H}$
can be chosen to be states with definite parity.
(The same is the case for ABJM theory, whereas for ABJ theory parity 
is broken at the non-planar level~\cite{Kristjansen:2008ib,Caputa:2009ug}.)
In general, for ${\cal N}=4$ SYM with gauge group $SU(N)$, for a given length $L$ the 
spectrum will then contain operators of positive as well as negative parity.
However, since the group generators for gauge group $SO(N)$ are antisymmetric, a state is 
related to its parity conjugate in the following way: 
\begin{equation}
\hat{P}\mbox{Tr}(X_{i_1} X_{i_2}\ldots X_{i_L})
=(-1)^L \mbox{Tr}(X_{i_1} X_{i_2}\ldots X_{i_L}).
\label{parityconjugate}
\end{equation}
In other words, parity has been gauged. We thus see that, compared to the case of 
$SU(N)$, the $SO(N)$ theory has a lot fewer states: For even length only
 positive parity states survive whereas for odd length only negative parity states survive. 
When acting on operators of the type~(\ref{operators}), the 
one-loop dilatation generator $\hat{H}$ can be usefully decomposed as
\begin{equation}
\hat{H}=N\,\hat{H}_0 +\hat{H}_++ \hat{H}_- + \hat{H}_{flip}.
\label{Hamexp} 
\end{equation}
Here $\hat{H}_0$ is the planar part which, up to a factor of two, is the same as for $SU(N)$, 
i.e.\footnote{The relative factor of $\frac{1}{2}$ in the hamiltonian arises because of our 
normalisation of the gauge group generators.}
\begin{equation} \label{Hnormalisation}
\hat{H}_0^{SO(N)}\equiv\hat{H}_0=\frac12\sum_{i=1}^L (1-P_{i,i+1})
=\frac{1}{2}\hat{H}_0^{SU(N)}.
\end{equation}
In particular, this means that the information about the planar anomalous dimensions in the case of 
gauge group  $SO(N)$ is encoded in the same Heisenberg spin chain Bethe equations as for $SU(N)$. 
However, due to the fact that certain states are projected out, some of the other information encoded 
in these equations becomes redundant. 

For single trace operators consisting of $M$ fields of type $\phi$ and $(L-M)$
fields of type $Z$, where $M\leq L/2$, the Bethe equations are expressed in terms of $M$ 
rapidities $\{u_k\}_{k=1}^M$ and  read
\begin{equation}
\left(\frac{u_k+\frac{i}{2}}{u_k-\frac{i}{2}}\right)^L=
\prod_{j=1,j \neq k}^M \frac{u_k-u_j+i}{u_k-u_j-i}.
\label{oneloopplanar}
\end{equation}
The rapidity $u$ is related to the momentum $p$ via
\begin{equation}
u=\frac{1}{2}\cot \left(\frac{p}{2}\right),
\end{equation}
and the eigenvalues of $\hat{H}_0$ are given by
\begin{equation}
E_0=\frac{1}{2}\sum_{k=1}^M \frac{1}{u_k^2+\frac{1}{4}}=2\sum_{k=1}^M \sin^2 
\left(\frac{p_k}{2}\right).
\end{equation}
The momenta have to satisfy the condition 
\begin{equation}
\sum_k p_k=0, \label{cyclicity}
\end{equation}
which reflects the cyclicity of the trace. The Bethe equations,
the cyclicity constraint and the expression for the energy are all invariant
under $u_k\rightarrow -u_k$. This implies that for 
any solution, $\{u_k\}$, either $\{-u_k\}=\{u_k\}$ or
$\{-u_k\}$ is a partner solution of the same energy. 
Following~\cite{Beisert:2003xu,Beisert:2004hm},
we will refer to the
first type of solutions as unpaired solutions and the second type as paired.
In $SU(N)$ terminology, the two solutions in a pair are each other's parity 
conjugates. The values of the higher conserved charges for the two states 
are identical for even charges and differ by a sign for odd charges. Unpaired 
states have vanishing odd charges.
Considering gauge group $SO(N)$ instead of $SU(N)$, the two
states in a pair get identified via eqn.~(\ref{parityconjugate})
and the odd charges lose their meaning. An unpaired state 
survives the projection if it has parity $(-1)^L$ where $L$ is its
length.
The reduction procedure is hence clear on the level of solutions. It
would be neat, however, if it could be formulated at the level of the Bethe 
equations.\footnote{One can show that the surviving unpaired states always
have $L$ and $M$ even~\cite{AIpsen}.
For these states, one can hence 
directly see that the Bethe equations will take a form like 
$$
\left(\frac{u_k+\frac{i}{2}}{u_k-\frac{i}{2}}\right)^{L-1}=
\prod_{j=1,j \neq k}^{M/2} \frac{u_k-u_j+i}{u_k-u_j-i}\frac{u_k+u_j+i}{u_k+u_j-i}
$$
which is similar to the (not completely unrelated) case of open 
strings \cite{Chen:2004mu,Chen:2004yf,Erler:2005nr,Beisert:2008cf}.}

\label{parityconstraint}

\begin{figure}[h]
\begin{center}
\rotatebox{270}{\includegraphics[width=3.0cm]{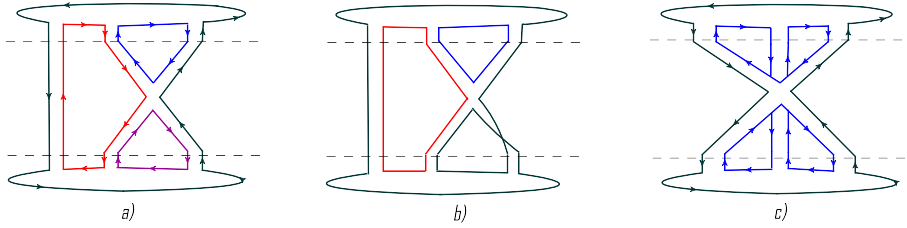}}
\end{center}
\caption{A planar Feynman diagram (a), a non-orientable
diagram with a single cross-cap (b) and
a diagram of genus one (c).}
\label{Nonorientable}
\end{figure}

At the non-planar level the dilatation operator contains the three terms
$\hat{H}_+$, $\hat{H}_-$ and $\hat{H}_{flip}$.
The operators $\hat{H}_+$ and $\hat{H}_-$ 
respectively increase and decrease the trace number by one and have analogues
in the case of $SU(N)$. The operator $\hat{H}_{flip}$ is trace conserving
and does not have any analogue in the case of $SU(N)$. 
In the language of string theory the operators
$\hat{H}_+$ and $\hat{H}_-$ 
correspond to string splitting and joining whereas $\hat{H}_{flip}$
corresponds to the insertion of a cross-cap on the string worldsheet.
It is well-known that for gauge theories with orthogonal or symplectic
gauge group the topological expansion includes Feynman diagrams which 
correspond to non-orientable surfaces, i.e.\ surfaces with 
cross-caps~\cite{Cicuta:1982fu}. Each occurrence
of a cross-cap is associated with a factor of $\frac{1}{N}$ whereas 
a handle as usual gives rise to a factor of $\frac{1}{N^2}$, see
Fig.~\ref{Nonorientable}.
Acting with  $\hat{H}_{flip}$ on a single trace operator 
gives a contribution for each
pair of fields of type $\phi$, $Z$ that the operator
contains. This contribution  is most conveniently
described in the following way
\begin{equation} \label{Hflip}
\hat{H}_{flip}\mbox{Tr}(\phi X Z Y) = 
\frac{1}{2}Tr( X^T Y [Z,\phi])
+\frac{1}{2}\mbox{Tr}(Y X^T [Z,\phi]).
\end{equation}
Here $X$ and $Y$ are arbitrary operators, and it is
understood that the $\check{Z}$ and $\check{\phi}$ in $\hat{H}_{flip}$
are contracted with the explicitly written $Z$ and $\phi$ in 
$\mbox{Tr}(\phi X Z Y)$.
The operator $\hat{H}_{flip}$ hence cuts out a piece of the operator
and reinserts it with the opposite orientation. Since this piece
can be of arbitrary length, we see that all sites in the chain are 
involved in the interaction. So, although $\hat{H}_{flip}$ takes single-trace operators 
to single-trace operators, and can thus be interpreted as a spin-chain interaction, in constrast
with the planar part of the dilatation operator 
its action on the spin chain is highly non-local.

Up to a factor of 2, the operator $\hat{H}_+$ takes the same form for
$SU(N)$ and $SO(N)$ whereas the operator $\hat{H}_-$ has extra terms for 
$SO(N)$. More precisely 
\begin{eqnarray}
\hat{H}_+^{SO(N)}&=&\frac{1}{2}\hat{H}_+^{SU(N)} \\
\hat{H}_-^{SO(N)} \mbox{Tr}(\phi X) \mbox{Tr} (Z Y)
&=&\frac{1}{2}\hat{H}_-^{SU(N)} \mbox{Tr}(\phi X) \mbox{Tr} (Z Y)  \nonumber \\
& & \mbox{}+\frac{1}{2}\mbox{Tr}(X^T Y [\phi, Z]  )+
\frac{1}{2}\mbox{Tr}(Y X^T [Z, \phi]),
\label{H-SO}
\end{eqnarray}
where the notation is as above and where $\hat{H}_\pm^{SU(N)}$ can be found
in~\cite{Beisert:2002ff}.
The extra terms in $\hat{H}_-^{SO(N)}$ are natural
since for non-orientable surfaces there are two possible ways of
gluing objects together. We notice that in a basis of planar 
eigenstates the perturbations $\hat{H}_+$ and
$\hat{H}_-$ are always off-diagonal. Only $\hat{H}_{flip}$
can have diagonal matrix elements in such a basis. Treating the energy
corrections perturbatively in $\frac{1}{N}$, $\hat{H}_+$ and $\hat{H}_-$ will thus 
generically give
corrections to the energy of order $\frac{1}{N^2}$ whereas $\hat{H}_{flip}$
can give corrections already at order $\frac{1}{N}$. The expansion of the
anomalous dimensions hence generically takes the form
\begin{equation}
{\cal E}= \frac{g_{\mbox{\tiny YM}}^2 N}{8\pi^2}
\left( E_0+\frac{1}{N}E_1+\frac{1}{N^2}E_2+{\cal O}
\left(\frac{1}{N^3}\right)\right),
\end{equation}
where the contribution $E_1$ is mainly due to $\hat{H}_{flip}$.
It should be noticed, however, that if there are degeneracies in the planar spectrum, energy 
corrections induced by $\hat{H}_+$ and $\hat{H}_-$ can also be of order $\frac{1}{N}$. This
phenomenon does not occur for strong coupling where the closed string
perturbation theory taking into account string splitting and
joining always gives rise to an expansion in $\frac{1}{N^2}$. The 
$\frac{1}{N}$ corrections to the energies induced by $\hat{H}_{+}$ and 
$\hat{H}_-$
are hence expected to vanish for strong coupling (and only arise here
due to an order of limits issue). Assuming this to be true we can thus
study corrections to the string energy induced by cross-caps by considering
only the corrections coming from $\hat{H}_{flip}$. 

\section{${\cal N}=4$ SYM with gauge group $Sp(N)$.}

We now consider the case of $\Ncal=4$ SYM with gauge group $Sp(N)$, the group of $N\times N$ symplectic
matrices. The construction of this theory in terms of an orientifold projection is also well known \cite{Gimon:1996rq}:
The projection in this case relates the Chan-Paton matrices of open-string states as 
\be
\lambda=-J^{-1}\lambda^T J
\ee 
where $J$ is an antisymmetric matrix satisfying $J^{\,2}=-1_{N\times N}$ , which can be taken to be ($N$ is even):
\begin{equation}
J= \left(\begin{array}{cc}
0 & 1\\
-1
&  0
\end{array} \right)_{N\times N}.
\end{equation}
The Chan-Paton matrices in this case turn out to be symmetric, and generate the adjoint representation 
of $Sp(N)$. Combining this with the identification $X^i\ra -X^i$ of the $\Ncal=4$ SYM scalars leads
to $\Ncal=4$ SYM theory with gauge group $Sp(N)$ \cite{Witten:1998xy}.

In $Sp(N)$, indices are raised and lowered with the matrix $J$, and  adjoint fields 
with both indices down are symmetric. Thus an adjoint field $Z^\alpha_{\;\beta}=J^{\alpha\gamma}Z_{\gamma\beta}$ 
behaves in the following way under transposition
\begin{equation}
Z^T= J Z J\;. \hspace{0.7cm} 
\end{equation}
This in particular implies that a single trace operator is again related to
its parity conjugate as given in eqn.~\rf{parityconjugate} and parity is
gauged in the same way as before.
Furthermore, for gauge group $Sp(N)$ the one-loop dilatation generator of
${\cal N}=4$ SYM can again formally be expressed in exactly the same form as for 
$SU(N)$, cf. eqn.~(\ref{dilop}). Only the contraction rules are different. More
precisely one has
\begin{equation}
\check{Z}_{\alpha\,\beta} Z_{\gamma\epsilon}
=\frac{1}{2}(\delta_{\alpha \epsilon} \delta_{\beta \gamma}-
J_{\alpha \gamma} J_{\beta \epsilon}).
\end{equation}
Again one finds that the Hamiltonian can be written in the form given
in~(\ref{Hamexp}).
The action of $\hat{H}_{flip}^{Sp(N)}$ can be presented in the following way
\begin{equation}
\hat{H}_{flip}^{Sp(N)}\mbox{Tr}(\phi X Z Y) = 
\frac{1}{2}\mbox{Tr}( J X^T J Y [Z,\phi])
+\frac{1}{2}\mbox{Tr}(Y J X^T J [Z,\phi]).
\end{equation}
We notice that the result differs from that of $SO(N)$ by $X^T$ being replaced
by $J X^T J$. This difference amounts to a shift of sign as we have for an
operator $X$ of length $L$
\begin{eqnarray}
SO(N): && \hspace{0.5cm} X^T = (-1)^L \, \hat{P} X, \\
Sp(N): && \hspace{0.5cm}J X^T J = (-1)^{L+1} \, \hat{P} X, 
\end{eqnarray}
where $\hat{P}$ is the parity operator.
This is in full accordance with the general result that $SO(N)$ can 
be understood as 
$Sp(-N)$~\cite{Mkrtchian:1981bb,Cvitanovic:1982bq}.
Notice that this sign difference need not explicitly manifest itself
in the off-diagonal terms $\hat{H_+}$ and $\hat{H}_-$ since these will
generically give rise to energy corrections of order $\frac{1}{N^2}$.
For $Sp(N)$ we again find that the operator $\hat{H}_+$ differs
from that of $SU(N)$ only by a factor of $\frac{1}{2}$ whereas the
operator $\hat{H}_-$ has extra terms compared to the corresponding operator
for $SU(N)$. More precisely
\begin{eqnarray}
\hat{H}_+^{Sp(N)}&=& \frac{1}{2}H_+^{SU(N)} \\
\hat{H}_-^{Sp(N)} \mbox{Tr}(\phi X) \mbox{Tr} (Z Y)
&=&\frac{1}{2}\hat{H}_-^{SU(N)} \mbox{Tr}(\phi X) \mbox{Tr} (Z Y)   \\
& &+\frac{1}{2}\mbox{Tr}(JX^T J Y [\phi, Z]  )+
\frac{1}{2}\mbox{Tr}(Y JX^TJ [Z, \phi]).
\nonumber
\end{eqnarray}
The difference between the extra terms for
$Sp(N)$ and $SO(N)$ is that $X^T$ is
replaced by $J X^T J$, cf.\ eqn~(\ref{H-SO}), which as before amounts
to a change of sign. 

\section{Analysis of BMN operators \label{BMN}}
BMN operators are operators consisting of a background of $Z$ fields and
a finite number of excitations in the form of $\phi$-fields. We will restrict
ourselves to discussing the simplest operators of this type, i.e.\ those
having two excitations. Two-excitation BMN operators always have positive 
parity and therefore in the case of gauge group $SO(N)$  exist only for
even length. At the planar level a basis for the
two-excitation states can be chosen as
\begin{equation}
O_p^J=\mbox{Tr}(\phi Z^p \phi Z^{J-p}), \hspace{0.7cm} 0\leq p\leq J.
\end{equation}
In terms of these the eigenstates of $\hat{H_0}$ read
\begin{equation}
|n\rangle\equiv {\cal O}_n^{J}=\frac{1}{J+1}\sum_{p=0}^J\cos 
\left(\frac{\pi n(2p+1)}{J+1}\right){\cal O}_p^J,
\hspace{0.7cm}0\leq n\leq \frac{J}{2},
\end{equation}
and the corresponding eigenvalues are
\begin{equation}
E_0^n= 4 \sin^2\left( \frac{\pi n}{J+1}\right). 
\end{equation}
The inverse transformation giving ${\cal O}_p^J$ in terms of 
$|n\rangle$ takes the form
\begin{equation}
{\cal O}_p^J= |0\rangle+2
\sum_{n=1}^{J/2} \cos\left( \frac{\pi n(2p+1)}{J+1}\right)
|n\rangle.
\end{equation}
The energy correction induced by the perturbation
$\hat{H}_{flip}$ is simply given by the expression from first order quantum
mechanical perturbation theory, i.e.
\begin{equation}
E_1^n=\langle n| \hat{H}_{flip}|n\rangle.
\end{equation}
In order to determine this quantity we first evaluate 
$\hat{H}_{flip}{\cal O}_p^J$ where $J$ is assumed to be even. We find
(after some manipulations)
\begin{eqnarray}
\hat{H}_{flip} {\cal O}_p^J&=&
-\frac{1}{4}
(1-(-1)^p)\left\{2{\cal O}_p^J-{\cal O}_{p-1}^J-{\cal O}_{p+1}^J\right\}
\nonumber \\
&&
-\frac{1}{2}(-1)^p\left\{ {\cal O}_0^J+{\cal O}_J^J+
2\sum_{k=1}^{J-1}(-1)^k {\cal O}_k^J\right \}.
\end{eqnarray}
Having this expression, it is straightforward to determine the general
matrix element of $\hat{H}_{flip}$ as all sums involved are geometric sums.
The result reads
\begin{eqnarray}
\lefteqn{\langle m| \hat{H}_{flip}| n \rangle = }\nonumber\\
&\mbox{ }&-\frac{1}{J+1}\sin^2 \left(\frac{\pi m}{J+1}\right)
\left\{\delta_{n,m}(J+1)-\frac{1}{\cos\left(\frac{\pi(n-m)}{J+1}\right)}
-\frac{1}{\cos\left(\frac{\pi(n+m)}{J+1}\right)}\right\}  \nonumber \\
& & -\frac{2}{J+1}\frac{\sin^2\left(\frac{\pi m}{J+1}\right)}
{\cos \left(\frac{\pi n}{J+1}\right) \cos\left(\frac{\pi m}{J+1}\right)}.
\label{generalelement}
\end{eqnarray}
We notice that $\hat{H}_{flip}$ is not hermitian but this phenomenon
is well-known~\cite{Janik:2002bd,Beisert:2002ff}:
The operator $\hat{H}_{flip}$ is
related to its hermitian conjugate by a similarity transformation.
For $n=m$ the expression~(\ref{generalelement}) reduces to
\begin{eqnarray}
E_1^n&=& \langle n |\hat{H}_{flip}| n \rangle  \\
&=& -\frac{2}{J+1}\tan^2\left(\frac{\pi n}{J+1}\right)-
\frac{1}{J+1}\sin^2 \left(\frac{\pi n}{J+1}\right)
\left(J-\frac{1}{\cos\left(\frac{2\pi n}{J+1}\right)}\right). \nonumber
\label{diagonalelement}
\end{eqnarray}
This should correspond to the energy correction to a closed string state
resulting from the insertion of a cross-cap on its worldsheet. 
Defining $\lambda'=g_{{\mbox{\tiny YM}}}^2 N/J^2$ and $g_2= J^2/N$, the 
anomalous
dimensions of BMN operators were
originally believed to have a double expansion in $\lambda'$ and $g_2$ in
the limit $\lambda, J, N\rightarrow \infty$ with $\lambda',g_2$ 
fixed~\cite{Berenstein:2002jq,Kristjansen:2002bb,Constable:2002hw}. 
This double expansion
worked for BMN operators in ${\cal N}=4$ SYM with gauge group $SU(N)$
for the first few terms in $\lambda'$ and $g_2$ and led to some success
in reproducing the first non-planar correction on the gauge theory side
from LCSFT, for a review see~\cite{Grignani:2006en}. Later 
it was understood
that planar BMN scaling breaks down at four loop order in the gauge 
theory~\cite{Eden:2006rx,Beisert:2006ez,Bern:2006ew}.
Furthermore, on the string theory side a BMN expansion would involve
half-integer powers of $\lambda'$ starting at
one-loop order~\cite{Beisert:2005cw}.
 Here the first few terms of the expansion in powers of $\lambda'$ and $g_2$ for
the anomalous dimension in eqn.~(\ref{diagonalelement}) read
\begin{equation}
{\cal E}^n = \frac{\lambda'}{2}\,\left( n^2-g_2\,\frac{ n^2}{4 J^2}\right),
\end{equation}
meaning that the first non-planar contribution would not survive the 
above mentioned limit. Still it would be interesting to analyse the
cross-cap scenario in the pp-wave geometry by some version
of LCSFT.

\section{Search for integrability at finite $N$ \label{integrability}}

For gauge group $SU(N)$ an important concept in the search for integrability
was the occurrence of so-called planar 
parity pairs, i.e.\ pairs of operators which at the planar level had the
same anomalous dimension but opposite parity. The existence of such parity 
pairs could be traced back to the existence of an extra conserved charge 
commuting with the Hamiltonian but anti-commuting with 
parity~\cite{Beisert:2003tq}. 
When splitting and joining of traces were taken into account the degeneracy between
the operators in a parity pair disappeared and this was taken as an indication
that integrability was lost beyond the planar level~\cite{Beisert:2003tq}.
The situation was the same for ABJM theory~\cite{Kristjansen:2008ib}. 
In the case of gauge group $SO(N)$ where parity is gauged one obviously
does not even have planar parity pairs. Thus one has to
invent other means to test for integrability. 

One option is to look for other types of degeneracies in the spectrum which
could survive the non-planar corrections. One such type of degeneracy is 
that between anomalous dimensions of certain
single- and multi-trace operators, for instance between  BMN operators with
different number of traces, i.e.\ operators of the type
\begin{equation}
{\cal O}_n^{J_0;J_1,\ldots J_k}\equiv {\cal O}_n^{J_0}\, 
\mbox{Tr}(Z^{J_1})\mbox{Tr}(Z^{J_2})\ldots \mbox{Tr}(Z^{J_k}),
\end{equation}
with anomalous dimension
\begin{equation}
E_{0;n}^{J_0;J_1,\ldots J_k}=4\sin^2(\frac{\pi n}{J_0+1}).
\end{equation}
These degeneracies between BMN states with different numbers of traces
were what rendered the non-planar problem of ${\cal N}=4$ SYM with gauge
group $SU(N)$ intractable. The degeneracies are less pronounced in the case
of gauge group $SO(N)$ due to the gauging of the parity symmetry.
The first case of planar degenerate BMN states in the 
$SO(N)$ case is the degeneracy between the states ${\cal O}_3^8$ and
${\cal O}_1^{2;4}$.  The second case is the degeneracy between the operators
${\cal O}_5^{14}$ and ${\cal O}_3^{8;6}$. Using the full Hamiltonian
we can easily check if the first non-planar
correction which is of order $\frac{1}{N}$ lifts the degeneracy in these
two cases and it turns out that it does. There is thus no hint of non-planar
integrability from this analysis.

Another option to test for integrability is to directly try to construct
conserved charges commuting with the Hamiltonian. 
In the higher loop analysis of ${\cal N}=4$ SYM it was found that
such conserved charges could be constructed order by order in the 
coupling constant, $\lambda$~~\cite{Beisert:2003tq}. More generally one can
generate perturbatively integrable long range spin 
chains with GL(K) symmetry starting from
chains with nearest neighbour 
interactions~\cite{Beisert:2005wv,Beisert:2008cf}.
The construction can be elegantly described in terms of a master
symmetry~\cite{Fokas:1981cd} or a boost operator~\cite{Tetelman:1982}
and leads to a large family of long range perturbatively integrable spin 
chains~\cite{Bargheer:2008jt,Bargheer:2009xy}. 
These techniques do unfortunately not
immediately apply 
to our case as they require that the spin chain length exceeds the range
of the interaction. Nevertheless, we will discuss the possibility of 
constructing higher conserved charges perturbatively in $\frac{1}{N}$. 
For spin chains with local interactions integrability follows as soon as
a single additional charge commuting with the Hamiltonian can
be found~\cite{Grabowski:1994rb,Beisert:2007jv}. Again, this does not
necessarily apply to our type of spin chain.

Since, as discussed earlier, 
the odd charges lose their meaning in our setting, where parity is
gauged, at planar level the next higher conserved charge after the hamiltonian 
$\hat{H}=\hat{Q}_2$ is the even charge $\hat{Q}_4$. If we expand to first order
in $1/N$,
\begin{equation}
\hat{H}=  \hat{H}_0+ \frac{1}{N}\hat{H}_{flip},
\hspace{0.7cm}
\hat{Q}_4=\hat{Q}_4^{(0)}+\frac{1}{N} \hat{Q}_4^{(1)},
\end{equation}
our task is to determine a suitable $\hat{Q}_4^{(1)}$ such that
\begin{equation} 
[\hat{H}_0,\hat{Q}_4^{(1)}]+[\hat{H}_{flip},\hat{Q}_4^{(0)}]=0.
\label{commutation}
\end{equation}
Since $\hat{H}_{flip}$ only acts within a single trace, we can assume
the same about $\hat{Q}_4^{(1)}$. 
At the planar level, the higher charges can be constructed iteratively
starting from the Hamiltonian by means of the boost operator 
$\hat{B}$~\cite{Grabowski:1994ae}, i.e.\ 
\begin{equation}
[\hat{B},\hat{Q}_n^{(0)}]= \hat{Q}_{n+1}^{(0)},
\end{equation}
where $\hat{B}$ is a moment of the Hamiltonian:
\begin{equation}
\hat{B}=\frac{1}{2i}\sum_{j=1}^L j\, {\bf{\sigma}}_j\cdot {\bf{\sigma}}_{j+1},
\end{equation}
with the $\sigma$'s being the Pauli matrices.

Ignoring constants and terms commuting with $\hat{H}^{(0)}$, this gives\footnote{This matches the expression for $\hat{Q}_4^{(0)}$ given in \cite{Serban:2004jf}, up to the terms mentioned.} 
\be \label{Q4planar}
\hat{Q}_4^{(0)}=\sum_{i=1}^L\left(-8\left[P_{i,i+3}P_{i+1,i+2}-P_{i,i+2}P_{i+1,i+3}\right]+4P_{i,i+3}-4P_{i,i+2}\right)\;.
\ee
Lacking a constructive way of extending this expression beyond the planar level, we have tried to guess
a possible form by first rewriting all the permutation operators in terms of nearest-neighbour ones:
\be \label{permops}
\begin{split}
P_{i,i+3}&=P_{i+2,i+3}P_{i+1,i+2}P_{i,i+1}P_{i+1,i+2}P_{i+2,i+3} \quad\text{and}\;\;\\
P_{i,i+2}&=P_{i+1,i+2}P_{i,i+1}P_{i+1,i+2}
\end{split}
\ee
and then using the relation $P_{i,i+1}=I_{i,i+1}-2 H^{(0)}_{i,i+1}$ (cf. eqn. (\ref{Hnormalisation})) to rewrite $\hat{Q}_4^{(0)}$
in terms of the planar Hamiltonian. Having done this (with the caveat that the rewritings in (\ref{permops}) are 
not unique), it is then natural to introduce
a dependence on $\hat{H}^{flip}$ by perturbing as:
\be \label{defH0}
H_{i,i+1}^{(0)}\rightarrow H_{i,i+1}^{(0)}+\frac1N H^{flip}_i,
\ee
where we have decomposed  $\hat{H}^{flip}$ as 
\be
\hat{H}_{flip}=\sum_{i=1}^L H^{flip}_i\;.
\ee
More precisely, we define $H_i^{flip}$ by 
\be
H^{flip}_i=\sum_{j=1}^L H^{flip}_{ij} \;,
\ee
with $H^{flip}_{ij}$ acting on sites $i$ and $j$ of a {\it periodic} 
chain of length $L$ as, (cf. eqn. (\ref{Hflip})) \footnote{Note that there is an ambiguity in the location of the 
index $i$ on the chain after the action
of $H^{flip}_i$, which we have fixed by cyclically shifting the resulting chain by a suitable number
of sites, such that the first term of the commutator $[a_i,b_j]$ always ends up at position $i$. Keeping
track of $i$ is important when deforming the higher charges, since in a typical term $H^{flip}_i$ 
will be  preceded or followed by e.g. $H^{(0)}_{i,i+1}$ or $H^{(0)}_{i+1,i+2}$  and the sum over 
$i$ is performed only at the end.}
\be
\begin{split}
H^{flip}_{ij}& \left(\Mcal_{L-j+1,L-j+i-1}\otimes a_i\otimes \Ncal_{1,j-i-1} \otimes b_j \otimes \Mcal_{1,L-j}\right)\\
=&-\half \left((\Ncal^T\otimes \Mcal)_{L-i,L-2}\otimes [a_i, b_j]_\otimes\otimes (\Ncal^T\otimes \Mcal)_{1,L-i-1}\right)\\
&-\half \left((\Mcal\otimes \Ncal^T)_{L-i,L-2}\otimes [a_i , b_j]_\otimes\otimes (\Mcal\otimes \Ncal^T)_{1,L-i-1}\right)\;.
\end{split}
\ee
Here we have defined $\Mcal_{k,l}=m_k \otimes m_{k+1}\cdots m_{l-1}\otimes  m_{l}$ and similarly for $\Ncal$.  

The expression for $\hat{Q}_4^{(1)}$ obtained by inserting (\ref{defH0}) into (\ref{Q4planar}) is too long
to be reproduced here, but with the help of computer algebra we can check whether (\ref{commutation}) 
is satisfied. This turns out not to be the case for our naive guess for $\hat{Q}_4^{(1)}$. Given 
the amount of ambiguity involved in obtaining $\hat{Q}_4^{(1)}$, this is perhaps not surprising, 
and outlines the need for a more systematic approach.

A third way to look for integrability is to see if the first few non-planar
corrections can be reproduced from a perturbative Bethe ansatz as was the
case in the higher loop analysis of~\cite{Beisert:2003tq,Beisert:2004hm}. 
The most obvious way to check this is to simply try and derive a set of
Bethe equations, for instance using the coordinate space approach. This
direct approach is, however, not straightforward. First, it is not
clear how to implement the gauging of parity in a convenient way in this
language. Secondly, it is obvious that our spin chain does not have
an asymptotic regime since, as soon as we go beyond the planar
limit, all sites of the chain interact with each other. Therefore, we will
take a more naive approach.

Let us recall the perturbative Bethe equation for ${\cal N}=4$ SYM with gauge group
$SU(N)$. For operators of length $L$ containing $M$ $\phi$-fields and
$(L-M)$ $Z$-fields (with $M\leq L/2$) it reads
\begin{equation}\label{Bethex}
\left(\frac{x(u_k+\frac{i}{2})}{x(u_k-\frac{i}{2})}\right)^L
=\prod_{j\neq k}^M \frac{u_k-u_j+i}{u_k-u_j-i},
\end{equation}
where 
\begin{equation}
x(u)=\frac{1}{2}u+\frac{1}{2}\sqrt{u^2-2g^2} \equiv u(1-g^2f(u)),
\end{equation}
and where $g^2=\frac{g_{\mbox{\tiny YM}}^2 N}{8\pi^2}$.
Here $u$ is related to the momentum $p$ via
\begin{equation}
e^{ip}=\frac{x^+(u)}{x^-(u)},
\end{equation}
with
\begin{equation}
x^\pm(u)=x(u\pm \frac{i}{2}).
\end{equation}
For later convenience we notice that purely algebraic
arguments pertaining to the symmetry properties of the
full ${\cal N}=4$ SYM (and not just its $SU(2)$-sector) 
imply that one needs~\cite{Beisert:2005tm}
\begin{equation}
x^+ +\frac{g^2}{2x^+}-x^--\frac{g^2}{2x^-}=i,
\label{xpm}
\end{equation}
which is of course fulfilled by the function $x(u)$ given above.
Furthermore, we have the cyclicity constraint~(\ref{cyclicity})
and the energy is given as 
\begin{equation}
\label{dispersion}
E=\sum_k \frac{1}{g^2}\left(\sqrt{1+8g^2\sin^2(\frac{p_k}{2})}-1\right).
\end{equation}
For BMN states with two excitations we have $M=2$, $L=J+2$.
Following~\cite{Beisert:2004hm} and
expanding the Bethe root $u\equiv u_1=-u_2$ as
\begin{equation}
u=u_0+g^2 \delta u,
\end{equation}
we find from the Bethe equation to order $g^2$
\begin{equation}
\delta u=\frac{u_0}{u_0^2+\frac{1}{4}}\left(\frac{J+2}{J+1}\right),
\end{equation}
and consequently, with $E=E_0+g^2 \delta E$,
\begin{equation}
\delta E_{SU(N)}=
-16 \sin^4\left(\frac{n\pi}{J+1}\right)-64 \frac{1}{J+1} 
\cos^2\left(\frac{n\pi}{J+1}\right) \sin^4 \left(\frac{n\pi}{J+1}\right),
\end{equation}
where the first term comes from the correction to the dispersion
relation and the second one from the correction of the momenta.
Let us rewrite the first $\frac{1}{N}$-correction to the BMN states
of the $SO(N)$ gauge theory 
in a similar way
\begin{eqnarray}
\delta E_{SO(N)}&=&-\sin^2\left(\frac{n\pi}{J+1}\right) 
\label{energycorrection}\\
&&-\frac{1}{J+1}\left\{2\tan^2\left(\frac{\pi n}{J+1}\right)
-\frac{1}{2}
\tan^2\left(\frac{2 \pi n}{J+1}\right)\, \cos\left(\frac{2\pi n}{J+1}
\right)
\right\}. \nonumber
\end{eqnarray}
{}From this expression it is clear that if this were to arise from a Bethe
system the first term would have to originate from a correction of the
dispersion relation and the second one from a correction of the rapidities,
i.e.\ a correction of the Bethe equations. The needed correction of 
the rapidities would be
\begin{equation}
\delta u= -\frac{1}{J+1} \,\frac{4u_0^2+1}{64 u_0^3 \,(4u_0^2-1)}.
\label{deltau}
\end{equation}
There are of course many possible ways to deform the Bethe equations so that we
would get the rapidity corrections for two-excitation states appearing
in~(\ref{deltau}). Given a plausible deformation
one can test if it gives the correct answer for the energy of
states with more excitations which we can of course again compute using
quantum mechanical perturbation theory.
Let us illustrate this with a simple example. Parametrising the function
$x(u)$ as 
\begin{equation}
x(u)=u(1-\frac{1}{N} f(u)), \label{x(u)}
\end{equation}
we find that in order to correctly reproduce the
$\frac{1}{N}$-correction to the energies of the
two-excitation states the function $f(u)$ needs to fulfill the following equation
\begin{equation}
f_-(u)\equiv f(u+\frac{i}{2})-f(u-\frac{i}{2})=
-i \frac{1}{16u^3(4u^2-1)}.
\label{feqn}
\end{equation}
This implies that $f(u)$ can neither be written as a Taylor expansion
nor as a Laurent expansion in $u$. Notice, however, that to solve the modified
Bethe equations perturbatively we would only need to know $f_-(u)$. 
We have checked whether the Bethe equations with the expression for the 
$x(u)$ given in eqn.~(\ref{x(u)}) and the dispersion relation corrected by the
first term in~eqn.~(\ref{energycorrection}) correctly reproduce the energy of states with
four excitations and length eight, cf.\ Appendix~\ref{numerical}.
We found that the simple modification of the Bethe ansatz described above
does not lead to the correct non-planar correction
to the energy of any of these states. Now, one may ask whether the 
algebraic arguments which led to~(\ref{xpm}) and~(\ref{dispersion})
are valid for the non-planar case as well. I follows from the
analysis of reference~\cite{Beisert:2005tm} that the dispersion
relation can indeed be modified to include a correction which would
lead to the first term in the relation~(\ref{energycorrection}). 
However, the relation~(\ref{xpm}) to leading order in $\lambda$
simply becomes $x^+(u)-x^-(u)=i$ which
leads to the following constraint on the function $f(u)$
\begin{equation}
f(u+\frac{i}{2})+f(u-\frac{i}{2})=2iu \left[
f(u+\frac{i}{2})-f(u-\frac{i}{2})\right].
\label{constraint}
\end{equation}
This constraint is unfortunately incompatible with the 
relation~(\ref{feqn}).
Thus the naive proposal for the 
modification of the Bethe ansatz would anyway not have a chance to work
for the full ${\cal N}=4$ SYM theory.

Obviously, there are many
other possible ways to deform the Bethe ansatz. In particular, there is
the possibility of including a phase factor~\cite{Arutyunov:2004vx}.
This would, in the simplest possible approach, mean modifying the Bethe ansatz to
\begin{equation}
\left(\frac{u_k+\frac{i}{2}}{u_k-\frac{i}{2}}\right)^L
=\prod_{j\neq k}^M \frac{u_k-u_j+i}{u_k-u_j-i}
\left(1+\frac{i}{N}\, h(u_k-u_j)\right).
\label{mod2}
\end{equation}
Here we have for simplicity assumed that the phase factor depends only
on the difference of rapidities and that the modification of the Bethe
equations is due to the appearance of a phase factor alone. Demanding
again the modification of rapidities to be given by~(\ref{deltau}) we
find for the function $h(u)$
\begin{equation}
h(u)=\frac{1}{2 u^3\, (u^2-1)\,}. \label{heqn}
\end{equation}
Note the non-trivial fact that $h(u)$ is real for real $u$ and that 
$h(u)$ does not depend on the length of the spin chain. 
We have checked if the modified Bethe equation~(\ref{mod2}) correctly
reproduce the energy correction for length eight and four excitations.
Unfortunately, this is not the case. Needless to say that the tests
performed here do not exclude the existence of a modified Bethe ansatz.

\section{Comments on the string theory side \label{stringtheory}}

 As discussed in the previous sections, the spectral problem of $SO(N)$ and $Sp(N)$
$\Ncal=4$ SYM theory exhibits several interesting differences compared to 
the $SU(N)$ case. 
In this section we make some preliminary observations on how these differences manifest
themselves on the string theory side. 

 In sections 2 and 3 we sketched how the $\Ncal=4$ SYM theory with orthogonal or symplectic gauge
group can be obtained by performing an orientifold operation on a stack of D3-branes. Taking the 
near-horizon limit we find that the AdS/CFT dual gravity background should be given by an orientifold
of $\AdS_5\times\Srm^5$ \cite{Witten:1998xy}. Embedding the sphere in $\mathbb{R}^6$ as 
\be
\sum_{i=1}^6 (X^i)^2=1 \;,
\ee
this orientifold is a combination
of the $\Zset_2$ action $X^i\ra -X^i$ and the worldsheet orientation
reversal $\sigma\ra 2\pi-\sigma$. Note that the $\Zset_2$ acts without fixed points on $\Srm^5$ and thus
there is no orientifold plane. Consequently, there is no need for additional branes to cancel the 
orientifold plane charge, and thus no open string sector. Therefore, this setting still corresponds to an $\Ncal=4$
theory.\footnote{Orientifolds of $\Ncal=4$ SYM with fixed planes, which 
lead to $\Ncal=2$ conformal theories with additional flavours, have been considered in an integrability
context in \cite{Stefanski:2003qr,Chen:2004mu,Chen:2004yf}.} The dual geometry is now $\AdS_5\times \RP^5$,
and the difference between the $SO(N)$ and $Sp(N)$ projections lies in the presence of an additional B-field. 

 As discussed in \cite{Kakushadze:1998tr}, in the strict planar (free string) limit all correlation function
calculations in the orientifolded theory can be reduced, up to trivial rescalings, to those in the 
oriented one. We thus do not expect our picture of planar integrability to be modified in a major way. 
Of course, any spinning string solutions on $\Srm^5$ not invariant under the orientifold procedure will be 
projected out.

 Therefore, in the planar limit the differences to the $\Srm^5$ case are relatively minor and
arise only because  some spinning string 
solutions on $\Srm^5$ are not invariant under the orientifold transformation and are projected out. 
This corresponds to the fact, discussed in section 2, that certain gauge theory operators are projected
out, depending on their length and parity. Unfortunately, since the semi-classical string solutions
have large length, the distinction between odd and even length is not as 
apparent as on the gauge theory side. It would be interesting to do a thorough analysis of spinning
strings on $\AdS_5\times \mathbb{R}\mbox{P}^5$ along the lines 
of~\cite{Gubser:2002tv,Frolov:2002av,Frolov:2003qc} and we hope to return to this problem in the future. 

\begin{figure} 
\begin{center}
\includegraphics[height=4cm]{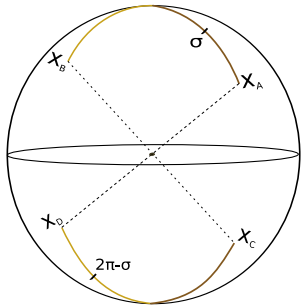}
\end{center}
\caption{A closed string solution on $\RP^2$ which is invariant under the orientifold. The configuration
$X(\sigma=0)=x_A$, $X(\sigma=\pi)=x_B=x_C\sim -x_B$, $X(\sigma=2\pi)=x_D$ is invariant under 
$X^i\rightarrow -X^i$ and $\sigma\rightarrow 2\pi-\sigma$.} \label{2strings}
\end{figure}

For the moment, however, we will confine ourselves to the straightforward observation that, by analogy with other contexts 
involving orientifolds, one can obtain invariant solutions by extending known ones with the addition of mirror strings. 
Let us demonstrate this for the $SU(2)$ sector, in which classical string solutions can be described in
terms of their profile on an $\Srm^2$ inside $\Srm^5$. This $\Srm^2$ is defined by $\sum_{i=1}^3 (x^i)^2=1$, where we have written
the coordinates of $\Srm^5$ as $X_1\pm i X_{4}=x^1 \exp(\pm i\phi_1)$, etc. 
Then the orientifold projection can be
taken to act on the coordinates of this $\Srm^2$ as $x^i\rightarrow -x^i$, resulting in the real
projective space $\RP^2$. Now, given any string solution with a 
profile $x^i(\sigma)$ for $0\leq\sigma<2\pi$
on $\Srm^2$, we can construct a ``doubled'' solution on $\RP^2$ by taking the profile to be $x^i(\sigma)$ for $0\leq\sigma<\pi$
and $-x^i(\sigma)$ for $\pi\leq\sigma<2\pi$. See Fig. \ref{2strings} for a drawing of such a solution on $\RP^2$. Note that,
despite appearances, the string in the figure is a closed string, since antipodal points are identified on $\RP^2$. 
The energy of such strings is always quadratic in $x^i(\sigma)$, so it will be exactly the same as the solution on 
$\Srm^2$.\footnote{For the purpose of comparing with weak coupling results, it might thus be more 
appropriate to use a different normalisation of the $SU(N)$ and $SO(N)$ 
generators in the gauge theory, 
or alternatively rescale the length of the string before and after the orientifold.} 

Arguing in this way, it seems that any solution which in the original 
$\AdS_5\times S^5$ geometry is confined to a half $S^2$ (the fundamental domain of $\RP^2$) inside the $S^5$, 
can be extended to a solution in $\AdS_5 \times \mathbb{R}\mbox{P}^5$ by superimposing it with
its mirror under the transformation $X_i\rightarrow -X_i$ and
$\sigma\rightarrow 2\pi -\sigma$. This includes for instance the giant
magnon solution~\cite{Hofman:2006xt} and the folded spinning string 
solution~\cite{Gubser:2002tv}.\footnote{Giant magnon solutions on $\RP^2$ have previously 
appeared in the context of the $AdS_4\times \Cset \mathrm{P}^3$ dual of ABJM theory, where the $\RP^2$ in that 
context arises as a suitable subspace of $\Cset \mathrm{P}^3$
\cite{Gaiotto:2008cg,Grignani:2008is,Abbott:2008qd,Abbott:2009um}.
The main difference in our case is that, since we are dealing with an
orientifold, we additionally need to implement the
worldsheet identification $\sigma\ra 2\pi-\sigma$.}

 Things become more interesting when considering $\frac{1}{N}$-corrections, which correspond to turning on
string interactions. Recall that the analogue of a spin chain splitting--and--joining operation
is a process where a string decays into two strings, which later recombine, creating a worldsheet 
of genus one. Such processes are not well understood, even in the pp-wave 
geometry, the main
obstacle coming from the necessity of summing over the infinite number of intermediate states 
(see \cite{Grignani:2006en} for a discussion). A simple 
model for splitting and joining of semi-classical strings in $\AdS_5\times\Srm^5$ was presented
in \cite{Peeters:2004pt}. However, as discussed (in a simplified model) 
in \cite{Casteill:2007td}, semi-classical splitting--and-joining does not seem to capture all 
of the relevant physics. 

\begin{figure}
\begin{center}
\includegraphics[angle=270, width=10cm]{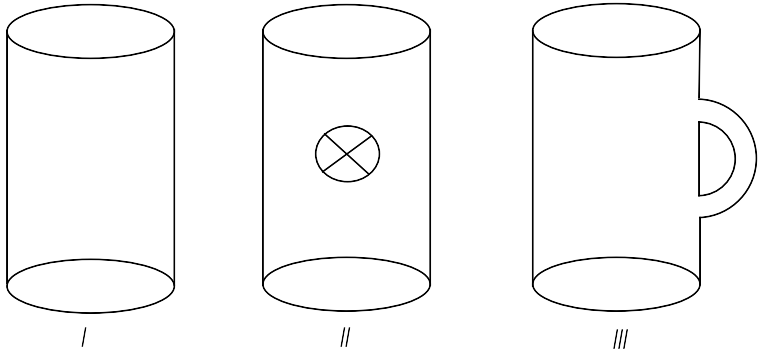}
\end{center}
\caption{Two-point string amplitudes. (I) The (planar) cylinder amplitude. (II) A cylinder with a cross-cap, contributing
at order $\frac{1}{N}$. (III) 
A cylinder with a handle, contributing at order $\frac{1}{N^2}$.}
\label{strdiag}
\end{figure}

 In our $SO(N)$ case, apart from the splitting--and--joining terms 
$\hat{H}_+$ and $\hat{H}_-$, the
dilatation operator contains an additional term which we have denoted by 
$\hat{H}_{flip}$. What is the
analogue of this term on the string side? It will clearly be related to the fact that, due to 
the orientifold operation, one should now also consider non-orientable string worldsheets, or in
other words worldsheets with cross-caps. Recall the weighting of a worldsheet 
with $b$ boundaries (each with $N$ Chan-Paton factors), $c$ cross-caps and $g$ handles:
\be
(N g_s)^b g_s^c g_s^{2g-2}=\lambda^{2g-2+b+c}N^{-c-2g+2},
\ee
where on the right-hand side we have rewritten the result in terms of gauge theory quantities, 
where the 't Hooft coupling is $\lambda=g_{YM}^2N=g_sN$. 
We see that a cross-cap weights the amplitude by a factor of $\frac{1}{N}$ 
compared to the oriented amplitude, while a handle by a factor of $\frac{1}{N^2}$. See Fig.~\ref{strdiag}. 
The cross-cap contribution thus, as expected, appears at the same 
order as the leading contribution from
$\hat{H}_{flip}$ on the gauge theory side and it is natural 
to identify the two. Intuitively, it is also clear that $\hat{H}_{flip}$
is associated with cross-caps since the operator acts by cutting out a
piece of an operator and gluing it back in with the opposite orientation.
Since it does not require summation over all intermediate states, the cross-cap calculation 
on the string theory side could be expected to be simpler than the genus-one case.

 It would be very interesting to perform such a non-oriented string calculation and compare with
the gauge theory side. Especially using a pp-wave geometry 
one might be able to compare with our
gauge theory results for BMN operators, cf.\ section~\ref{BMN}.

\section{Conclusion \label{conclusion}}

We have studied a number of features which distinguish the spectral problem
of ${\cal N}=4$ SYM with gauge group $SO(N)$ or $Sp(N)$ from that of 
${\cal N}=4$ SYM with gauge group $SU(N)$. Of particular interest to us was the
difference in the leading non-planar corrections. For orthogonal and
symplectic gauge groups the leading non-planar corrections
define a novel type of spin chain interaction of  highly non-local
nature which cuts out  a piece of the chain
and re-inserts it with the opposite orientation. Unlike the case of gauge
group $SU(N)$, the leading non-planar corrections a priori could fit
into the standard framework of integrability. However,
the resulting spin chain did not show any signs of integrability when 
studied by usual methods. In particular, our attempts to describe
the diagonalization problem for $\hat{H}_{flip}$ by means of
a Bethe ansatz were unsuccessful. However, given that the spin chain described by 
this Hamiltonian seems to lack an asymptotic regime (since all sites of the chain 
are involved in the interaction) it could still be that integrability, if present, simply
cannot be formulated in terms of a Bethe ansatz.

 Just as ${\cal N}=4$ SYM with orthogonal or symplectic gauge group
is much less studied
than its $SU(N)$ cousin, the same holds for the dual string theories. Here
we briefly discussed some issues related to studying the spectrum of
type IIB string theory on the
$\AdS_5 \times \mathbb{R}\mbox{P}^5$ background.
 We mentioned some features of spinning string solutions
and discussed how the leading non-planar corrections to anomalous dimensions
on the gauge theory side should originate from non-oriented string worldsheets
with a single cross-cap. By considering such worldsheets, one might hope to 
reproduce the leading non-planar corrections for two-excitation states
that we found from the gauge theory side. More generally, as cross-caps might be 
easier to handle than higher genus surfaces, this might open new possibilities for 
comparing gauge and string theories beyond the planar limit.

\vspace*{0.7cm}

\noindent
{\bf Acknowledgments:}
We thank  T.\ Lukowski, J.\ Plefka, A.\ Tseytlin, A.\ Wereszczynski, K.\ Zarembo
and especially N.\ Beisert for useful discussions. 
CK and KZ were supported by FNU through grant number 272-08-0329.
PC was supported in part by the Niels Bohr International Academy.

\appendix
\section{Numerical tests of Bethe equations.\label{numerical}}

We specify here the details of the numerical tests we performed. We focused
on the (single trace) states of length eight with four excitations. 
There are three such highest weight states. At one loop order at the
planar level they can 
be described in terms of the corresponding roots of the Bethe 
equations given in~\rf{oneloopplanar}. The three sets of roots $\{u_i^1\}$,
$\{u_i^2\}$ and $\{u_i^3\}$, $i\in\{1,2,3,4\}$
read\footnote{These roots as well as others
can be found in references~\cite{Beisert:2003xu,Beisert:2004hm}.}
\begin{eqnarray}
\{u_i^1\}&=& \{\pm 0.525,\pm 0.129\}, \\
\{u_i^2\}&=& \{\pm 0.0413,\pm 1.026i\}, \\
\{u_i^3\}&=& \{\pm0.463\pm 0.502 i\},
\end{eqnarray}
and the corresponding planar one-loop energies, $E_0^j$, $j=1,2,3$ are the roots of the
polynomial 
\begin{equation}
-x^3+10 x^2-29x +200=0.
\end{equation}
By direct diagonalization of $H_0+\frac{1}{N}\hat{H}_{flip}$ we find 
the $\frac{1}{N}$-corrections to the energies, $E_1^i$ to 
be\footnote{We remark that the operators considered here do not exhibit 
degeneracy with any multi-trace states and thus there are no further
corrections to their energies of order $\frac{1}{N}$.}
\begin{equation}
E_1^1= 1.618,\hspace{0.7cm}
E_1^2= -6.75, \hspace{0.7cm}
E_1^3= -19.85. \label{E1}
\end{equation}
On the other hand solving the Bethe ansatz~\rf{Bethex}
with $x(u)$ given by~\rf{x(u)}
and~\rf{feqn} we find the following $\frac{1}{N}$-correction to the 
rapidities 
\begin{eqnarray}
\delta u_1^i&=& \{\pm 0.0255\pm 0.000893i\},\\
\delta u_2^i&=& \{\pm 47.6,\pm 138.4i\},\\
\delta u_3^i&=& \{\pm 3.65,\pm 10.74\},
\end{eqnarray}
which leads to the following $\frac{1}{N}$-correction to the energies
\begin{equation}
E_1^1= -0.43, \hspace{0.7cm}
E_1^2=  -504,\hspace{0.7cm} 
E_1^3=  -26.6.
\end{equation}
These values clearly differ from the exact ones given in 
eqn.~\rf{E1}.

Using instead the deformed Bethe ansatz given by~\rf{mod2} and~\rf{heqn}
the $\frac{1}{N}-$ correction to the Bethe roots are
\begin{eqnarray}
\{\delta u_i^1\}&=&  \{\pm1.146\pm0.0327i\},\\
\{\delta u_i^2\}&=& \{\pm 5.96,\pm 17.29i\},\\
\{\delta u_i^3\}&=&  \{\pm 0.799,\pm 1.045\},
\end{eqnarray}
and the energy corrections, $E_1^i$ become
\begin{equation}
E_1^1= -2.07,\hspace{0.7cm}
E_1^2= -63.6,\hspace{0.7cm}
E_1^3= -8.25.
\end{equation}
These values also fail to agree with the exact ones given in 
eqn.~\rf{E1}.

\bibliography{SONfinal}
\bibliographystyle{JHEP}
\end{document}